# Doping-Induced Alterations in Electronic Structure of Copper Oxide Superconductors and a New Horizon for Higher Tc materials

Hiroshi Kamimura[1*], Jaw-Shen Tsai[1, 2], Osamu Sugino[3], Kunio Ishida[4], Hideki Ushio[5]

[1]*Tokyo University of Science, 1-3 Kagurazaka, Shinjuku-ku, Tokyo, 162-8601, Japan.*

[2]*RIKEN Center for Emergent Matter Science (CEMS), Wako, Saitama 351-0198, Japan*

[3]*University of Tokyo, Institute of Solid State Physics, 5-1-5 Kashiwanoha, Kashiwa, Chiba, 277-8581, Japan.*

[4]*Corporate Research and Development Center, Toshiba Corporation 1 Komukaitoshiba-cho, Saiwai-ku, Kawasaki 212-8582, Japan.*

[5]*Tokyo National College of Technology, Hachioji, 193-0997, Japan.*

By paying special attention to the fact that the doped holes induce deformation of $CuO_6$ octahedrons (or $CuO_5$ pyramids) in cuprate superconductors, we develop a non-rigid band theory treating doping-induced alterations of energy-band structures in copper oxide superconductors. Thanks to this theory, we obtain a complete picture of the doping-induced alteration in the electronic structure of $La_2CuO_4$, from the spin-disordered insulating phase to the metallic phase. We conclude that the Fermi surface structure of this cuprate in the underdoped region consists of Fermi pockets in the antinodal region and Fermi arcs in the nodal region, and thus that the origin of a so-called pseudogap is closely related to the existence of Fermi pockets. Moreover, we show that the carriers on the Fermi pockets contribute to the phonon mechanism in d-wave superconductivity. Finally, we discuss how one will be able to find higher Tc materials, based on the conclusions mentioned above.

Thirty years have passed since Bednorz and Müller discovered high temperature superconductivity in copper oxides (cuprates) by chemically doping hole carriers into $La_2CuO_4$, an antiferromagnetic (AF) insulator, by replacing 3+ cations ($La^{3+}$) by 2+ ones ($Ba^{2+}$) [1]. The $Cu^{2+}$ ion has a $d^9$ configuration, i.e. a one d-hole, in the parent material $La_2CuO_4$, so that one electron picture predicts the energy band as half-filled and the electronic structure as metallic, contradictory to the experimental result of the AF insulator. In this context, Anderson pointed out that a d hole is primarily localized at a Cu site to avoid strong Coulomb repulsion within a site, when the repulsion energy exceeds the kinetic energy gained by hopping to a neighboring Cu site [2]. The spins of these localized holes then form an AF order by the superexchange interactions via a closed-shell $O^{2-}$ ions between the neighboring Cu ions, resulting in a Mott insulator.

In this way, doped cuprates exhibit a fascinating class of electronic states from Mott insulators to d-wave superconductors, and moreover, show Tc reaching 135K in a family of mercury copper oxides [3]. Theoretical understanding, however, is far from complete, and in particular, a large gap exists between the itinerant electron picture and the molecular orbital (MO) picture, as explained below.

Starting from the hole carriers itinerating in a $CuO_2$ plane, a one-component theory was constructed using the t-J model [4]. Near the optimal doping condition, the theory yielded a metallic state with large Fermi surface (FS), which is in contradiction with the angle-resolved photoemission spectroscopy (ARPES) experiments. The experiments were then explained within this theory by an unconventional gap (a so-called pseudogap) opening near the antinodal region [5], but the mechanism leading to the pseudogap has not been fully settled [6].

Alternatively, starting from the hole carriers occupying MOs of a constituting $CuO_6$ octahedron unit, a ligand field theory was developed for the electronic structure of copper oxides [7]. According to this theory, the highest occupied band primarily consists of MOs degenerated in a regular octahedron with cubic symmetry. In addition, according to the Jahn-Teller (JT) theorem, the regular $CuO_6$ octahedron is unstable against deformation toward tetragonal symmetry such as elongation of the oxygen-Cu-oxygen axis perpendicular to the $CuO_2$ plane. The degeneracy is thus removed and the MO states split into two states, whose MOs are directing parallel to and perpendicular to the $CuO_2$ plane; the hole occupies the lower one. Taking into account these two MOs and their strong coupling to the JT distortion modes [8], a two-component model, called K-S model [9, 10, 11], was developed. This model is hardly solvable when the electron- electron interaction effects are incorporated, but a mean-field solution shows the appearance of small Fermi pockets. Nevertheless, the Fermi pocket was not observed at the predicted location in the ARPES [12].

In the previous mean-field treatment [12], the strong coupling with the JT distortion has been treated incompletely by taking the rigid band model. In this context, by a proper incorporation of the coupling with the JT distortion, we now achieved a consistent understanding of the electronic structures, successfully bridging the gap existing between the two different pictures.

In this paper, we present our theory by firstly showing an improved mean-field solution, which indicates a doping-induced alteration in the electronic structures of $La_{2-x}Sr_xCuO_4$ (LSCO), the simplest copper oxide superconductor, from a spin-disordered phase to a metallic phase. The result is then used to characterize the FS of the doped LSCO in terms of the Fermi pockets and the Fermi arcs, and the doping- and temperature- dependent behaviors of the Fermi pockets are used to elucidate the origin of the pseudogap. Finally, the holes in the Fermi pockets are shown to contribute to the phonon mechanism in the d-wave superconductivity.

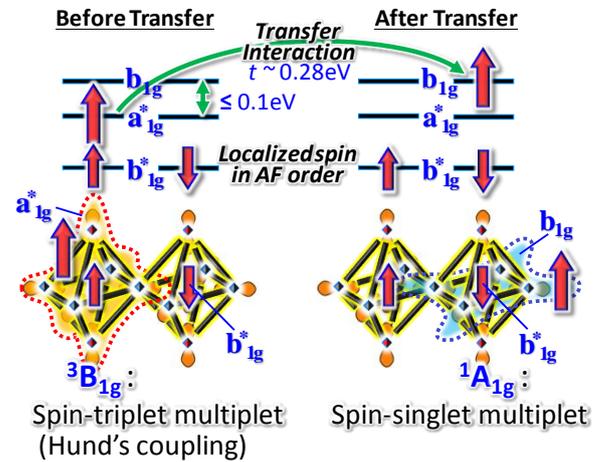

**Figure 1.** Transfer of a doped hole (a large arrow) from the Hund's coupling $^3B_{1g}$ to the spin-singlet $^1A_{1g}$ without destroying the AF order constructed by the $b^*_{1g}$ localized spins (small arrows). Here the $a^*_{1g}$ orbital consists of a Cu $dz^2$ orbital and surrounding six oxygen $p$ orbitals while the $b_{1g}$ orbital consists mainly of the four in-plane oxygen $p_\sigma$ orbitals hybridized by the Cu $dx^2$-$y^2$ orbital.

In the present paper, the two MOs in the JT deformed $CuO_6$ octahedron are called $a_{1g}$ antibonding orbital $|a_{1g}^*\rangle$ and $b_{1g}$ bonding orbital $|b_{1g}\rangle$, which are shown in Fig. 1. The AF order is originated from the superexchange interactions between the spin $S_i$ of a hole state called $b_{1g}$ antibonding orbitals $|b_{1g}^*\rangle$, which is deeply localized within each $i$-th octahedron. The unique exchange interaction appears between the localized spin $S_i$ and the spin of a doped hole, $s_{i,a_{1g}^*}$ or $s_{i,b_{1g}}$, within the $i$-th $CuO_6$ octahedron and is described by a term of the K-S Hamiltonian (see Eq. (1) of Ref. [13] for the full Hamiltonian):

$$H_{ex} = \sum_i \left( K_{a_{1g}^*} s_{i,a_{1g}^*} \cdot S_i + K_{b_{1g}} s_{i,b_{1g}} \cdot S_i \right). \quad (1)$$

Here $K_{a_{1g}^*}$ and $K_{b_{1g}}$, are, respectively, the exchange constants for the Hund's coupling spin-triplet state, $^3B_{1g}$, and for the spin-singlet state, $^1A_{1g}$. Their values were determined for the undoped LSCO as $K_{a_{1g}^*} = -2.0$ eV and $K_{b_{1g}} = 4.0$ eV [14]. By introducing the transfer interaction between $a_{1g}^*$ and $b_{1g}$ orbitals at neighboring sites, $t$ (= 0.28 eV) (see Eq. (1) of Ref. [13] for the full Hamiltonian), a coherent metallic state with local AF order is formed, by alternately combining $|a_{1g}^*\rangle$ and $|b_{1g}\rangle$ orbitals of the nearest neighbor octahedrons without destroying the AF order. This situation is schematically shown in Fig. 1. Thus, the key feature of the K-S model is the coexistence of a metallic state and local AF order.

Now we examine the effect of doping by replacing $La^{3+}$ by $Sr^{2+}$ in $La_2CuO_4$. We find that the JT deformation of the elongated $CuO_6$ octahedron is released partially, i.e. the two apical $O^{2-}$ ions approach towards a $Cu^{2+}$ ion at the center of the $CuO_6$ octahedron so as to gain the attractive electrostatic energy. This deformation which is opposite to the JT distortion is called the anti-Jahn-Teller effect [15]. By this effect, the magnitude of $K_{a_{1g}^*}$ increases owing to the associated increase in the overlap of $|a_{1g}^*\rangle$ with $|b_{1g}^*\rangle$ orbitals, while $K_{b_{1g}}$ remains unchanged because the $|b_{1g}\rangle$ orbital is extended on the $CuO_2$ plane. This effect was neglected previously by adopting the rigid band model [12].

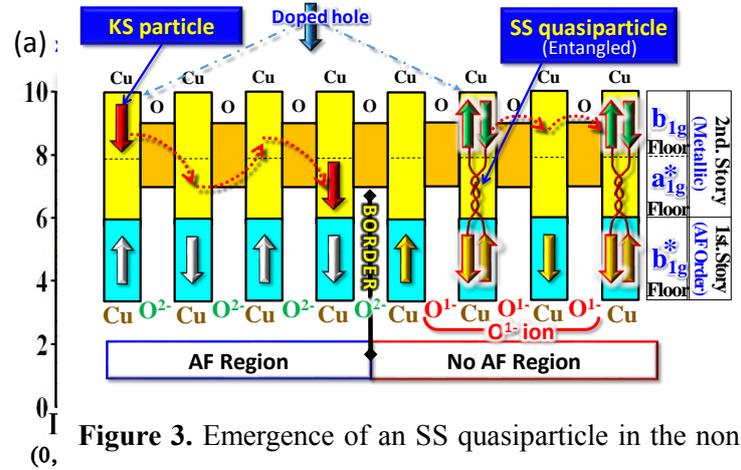

**Figure 3.** Emergence of an SS quasiparticle in the non-AF region (right side) and that of a KS particle in the AF region (left side). Here the K-S model is expressed schematically as a two-story house model (See Ref. [15]).

one $|K_{a*1g}(0)| = 2.0$ eV, arising from different overlaps between $b_{1g}^*$ and $a_{1g}^*$ orbitals (see inset).

Using the thus-improved non-rigid band theory, we calculated the energy band of the doped LSCO by replacing $S_i$ in Eq. (1) with their average value $\langle S \rangle$, as did for the rigid band model [12]. The resulting electronic structure exhibits a remarkable doping dependence. For the concentration $x = 0.03$ in the insulating phase below the experimental metal-insulator transition, $x_o = 0.05$ ($|K_{a*1g}| = 0.2 |K_{a*1g}(0)|$ with $|K_{a*1g}(0)| = 2.0$ eV), the top of the highest occupied band, #1 (see Fig. 2(a)), is located at the $\Delta$ point $(\pi/2a, \pi/2a)$ of a two-dimensional (2D) square Brillouin zone (BZ) (see

Fig. 4(a)). Here $a$ is the length of the Cu-O-Cu distance, which is taken as unity hereafter. The doped holes are thence accommodated in a region ranging from $(\pi/2, \pi/2)$ to a certain point on the side of the square BZ. Note that this square BZ (Fig. 4(a)) is a cross-section of 3D BZ (Fig. 4(d)) at the $k_x - k_y$ plane.

When the hole concentration increases above $x_o = 0.05$, the band structure becomes qualitatively different. Let us show in Fig. 4(b) the band structure calculated for $x = 0.10$ ($|K_{a*1g}| = 0.8|K_{a*1g}(0)|$), which is located between $x_o$ and the optimum doping $x_{opm} = 0.15$ in the underdoped regime. The state at the $G_1$ point $(\pi, 0)$ then becomes higher in energy than that at $(\pi/2, \pi/2)$, where the density of states shows a saddle-point singularity at $(\pi, 0)$. This alteration of the band shape has a particularly important meaning as discussed below.

This change in the top of the band induces a redistribution of doped holes, because the wave function at $(\pi/2, \pi/2)$ has a large population of an in-plane O $p_\sigma$ orbital, compared to that at $(\pi, 0)$. This means that the doping induces a charge-transfer forming $O^{1-}$ from $O^{2-}$, with an effect of weakening the superexchange interaction between the localized spins that requires a closed shell configuration of oxygen, i.e., $O^{2-}$. This charge-transfer thus assists a formation of non-AF regions. Thus, we call the insulating phase below $x_o = 0.05$ the spin-disordered insulating phase.

In such non-AF region, the localized spins orient randomly making the alternate arrangement of $|a^*_{1g}\rangle$ and $|b_{1g}\rangle$ unlikely to occur dominantly. The hole in $|b_{1g}\rangle$ will, instead, favor a formation of a spin-singlet (SS) coupling with the localized hole in $|b^*_{1g}\rangle$ through the exchange $K_{b_{1g}}$. The twisted lines between a $b_{1g}$ doped hole and a $b^*_{1g}$ localized hole within the same Cu site in the right side of Fig. 3 represent the entanglement state of an SS quasiparticle. The SS quasiparticle is known to favor the AF order because of the kinetic energy lowering mechanism [16-18], but the effect is not likely to recover the weakened superexchange completely, although it is possible that AF order is formed locally such that the SS quasiparticles can propagate coherently in a certain distance. Even in the presence of the local AF order, the SS quasiparticle will lose quantum coherence in the long run and will be affected by the Anderson localization.

According to the theories of Anderson [19] and Mott [20], a mobility edge $E_c$ exists such that those states with energy below (above) $E_c$ is localized (extended). This means that the electronic structure will change from the spin-disordered insulating phase to a metallic phase as the Fermi energy ($E_F$) of the SS quasiparticles crosses $E_c$ with increasing doping level. Since SS quasiparticles are constituted by the $b_{1g}$ hole states distributed around $(\pi/2, \pi/2)$ and the localized $b_{1g}^*$ hole states having no band dispersion, the corresponding FS structure is characterized by the distribution of $b_{1g}$ holes in the reciprocal space, which has a feature of the Fermi arc (see Fig. 4 for detail). Origin of the Fermi arc is thus explained by our theory differently from the one obtained from the t-J model [4, 21, 22].

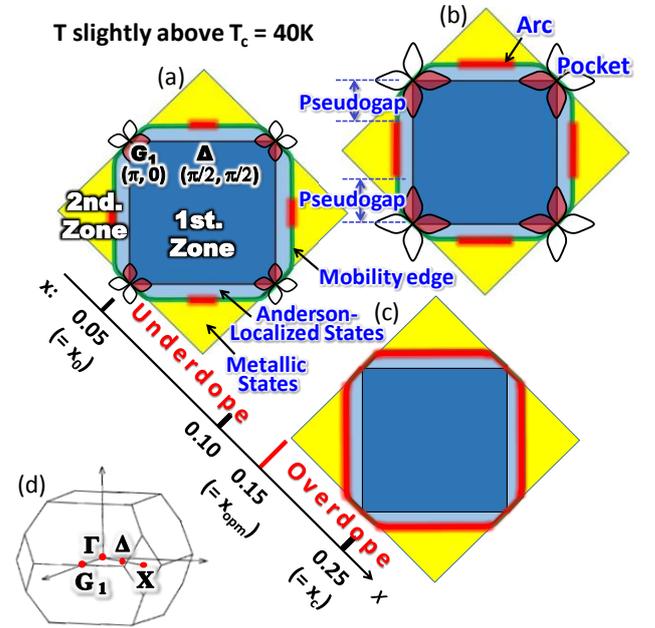

**Figure 4.** Fermi surface of LSCO and its doping dependence. Fermi arcs (red lines), Fermi pockets (red-white 4-pointed petal shapes) and mobility edge (green lines) for **(a)** x = 0.05, **(b)** $x = 0.10$, and **(c)** $x_c = 0.25$ shown in the $k_x - k_y$ plane of the Brillouin zone. **(d)** Ordinary Brillouin zone of LSCO with Γ, Δ, $G_1$ and X symmetry points.

At higher doping level above the metal-insulator transition, the top of the highest band switches from $(\pi/2, \pi/2)$ to $(\pi, 0)$. The wavefunction of $b_{1g}$ hole at $(\pi, 0)$ has a characteristic feature depicted by the K-S model. Namely, this hole forms a metallic state with the local AF order, by taking $|a_{1g}^*\rangle$ and $|b_{1g}\rangle$ orbitals alternately without destroying the AF order, as shown in the left side of Fig. 3. Thus, this unique hole is named as the KS particle. Considering the dynamical spin fluctuation typical of 2D spin systems, the FS constructed by KS particles coexisting with the local AF order will be observed experimentally in the AF BZ as a Fermi pocket at the $\Gamma$ point, (0, 0), where the AF BZ is obtained by folding the ordinary BZ in Fig. 4 (d).

From the calculated results of Figs. 2(a-b), and 3, we obtained the FS consisting of the Fermi pockets and Fermi arcs, respectively, in the antinodal and nodal region of the ordinary BZ. The results for $x = 0.05, 0.10$, and $0.25$ at a temperature just above $T_c$, shown in Figs. 4(a-c), exhibit a notable doping dependence. As $x$ increases from 0.05 to 0.15, the Fermi-arc areas expand together with the Fermi pockets with a 4-pointed petal shape, which is hole-like (red) and electron-like (white), respectively, inside and outside the second BZ. At $x = 0.25$, the Fermi pockets disappear completely and the Fermi arcs dominate. Therefore, $x = 0.25$ is named as the critical concentration $x_c$. We also find that the FS structure exhibits notable temperature dependence by the gradual destruction of local AF order by thermal agitation.

The calculated FS structure captures characteristics of the spectral weight maps at $E_F$ observed for $x = 0.07$ to 0.22 [23]. We expect our prediction (Figs. 4(a) to 4(c)) will be compared in detail with experiments to further advance the understanding of unusual behaviors of FS, which have been simply ascribed to the existence of pseudogap [6, 24, 25].

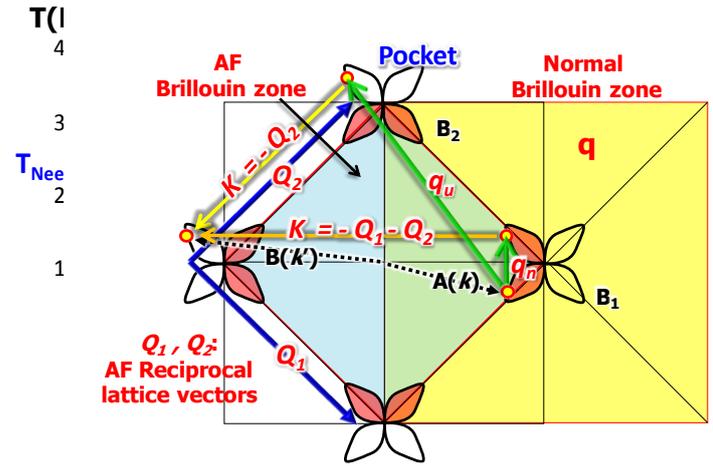

**Figure 6.** Verification of d-wave superconductivity. Illustration of two scattering processes by phonons: Normal and Umklapp.

In the following, we discuss an implication of the calculated results on the new phase diagram and superconductivity. Using the free-energy $F = E - TS$ calculated for the mixed phase consisting of KS particles and SS quasiparticles and for the pure SS phase consisting of SS quasiparticles only, where the internal energy $E$ and the electronic entropy $S$ are obtained from the calculated band structure and density of states of the highest #1 band [10, 11], we estimated the thermodynamic points of $T$ and $x$ at which both free-energies are equal. The resulting phase boundary between the mixed phase and the pure SS phase, $T^*(x)$, was used to construct the phase diagram (Fig. 5). Since we have previously shown that the Fermi pockets are responsible for the appearance of d-wave superconductivity in the phonon mechanism [13], the d-wave superconductivity should disappear at $= x_c$, at which $T^*$ vanishes. Thus, $T^*(x_c) = T_c(x_c) = 0$. The obtained phase diagram is consistent with the following experimental results: (1) Temperature-dependence of $T_c(x)$ [26]; (2) the opening of a symmetry gap only in the antinodal region in Bi2212 [27]; (3) the possible existence of antinodal quasiparticles [28]; and (4) the phase diagram for $Bi_2Sr_2CaCu_2O_{8+\delta}$ [29].

Occurrence of the d-wave superconductivity was previously discussed assuming the Fermi pockets to exist at the nodal region $(\pi/2, \pi/2)$ [10, 13, 30]. Although the present calculation shows that the Fermi pockets are located at the antinodal region $(\pi, 0)$, the difference in the location of Fermi pockets is not relevant because the Fermi pockets in both cases have the four-fold symmetry, which is an essential key to the d-wave superconductivity.

The wave functions of up and down spins in the K-S model, $\Psi_{k,\uparrow}(r)$ and $\Psi_{k,\downarrow}(r)$, have the unique phase relation expressed by $\Psi_{k,\downarrow}(r) = \exp(i\mathbf{k}\cdot\mathbf{a})\Psi_{k,\uparrow}(r)$, where $\mathbf{a}$ is a vector connecting Cu atoms along the Cu-O-Cu line, as focused previously [13], and is summarized below: The phonon with wave vector $\mathbf{q}$ scatters a KS particle at state $\mathbf{k}$ (at **A** point) to $\mathbf{k}'$ (at **B** point) through a spin-dependent interaction $V_\uparrow(\mathbf{k},\mathbf{k}',\mathbf{q}) = \exp(i\mathbf{K}\cdot\mathbf{a})V_\downarrow(\mathbf{k},\mathbf{k}',\mathbf{q})$, where $\mathbf{K} = \mathbf{k} - \mathbf{k}' + \mathbf{q}$ is a reciprocal lattice vector in the 2D AF BZ (see Fig. 6). The $\mathbf{K}$ satisfies either $\mathbf{K} = \mathbf{k} - \mathbf{k}' + \mathbf{q}_n = -\mathbf{Q}_1 - \mathbf{Q}_2$ (normal) or $\mathbf{K} = \mathbf{k} - \mathbf{k}' + \mathbf{q}_u = -\mathbf{Q}_2$ (Umklapp), corresponding respectively to $\exp(i\mathbf{K}\cdot\mathbf{a}) = +1$ (attractive interaction), or $\exp(i\mathbf{K}\cdot\mathbf{a}) = -1$ (repulsive interaction), where $\mathbf{Q}_1$ and $\mathbf{Q}_2$ are the AF reciprocal lattice vectors. When the Fermi pocket has four-fold symmetry, these scattering processes proceed coherently to yield a $d_{x^2-y^2}$-wave superconductivity [13]. In this consideration, Fermi arcs are not considered because they are formed by the SS quasiparticles in the non-AF region, and thus do not contribute to superconductivity.

At the end, we emphasize that electron-phonon interactions in cuprate are very strong. In Fig. 1(b), we have shown that two types of exchange interactions led to the emergence of the $^3B_{1g}$ and $^1A_{1g}$ multiplets. When these multiplets interact with one of the JT vibrational modes, say $A_{1g}$ out-of-plane mode, in which two apical oxygens oscillate perpendicular to the $CuO_2$ plane, the lowest state energies of these multiplets, $E(^3B_{1g})$ and $E(^1A_{1g})$, change with respect to variation of the Cu – apical O distance, c. According to Ref. [13], the energies depend on c as $dE(^1A_{1g})/dc = 2.2$ eVÅ$^{-1}$ and $dE(^3B_{1g})/dc = 2.8$ eVÅ$^{-1}$. These large values make remarkable contributions to the increase of the electron-phonon coupling constant for d-wave paring, $\lambda_d$, whose value is 1.96 for optimum doping ($x_{opm} = 0.15$) [13], indicating that cuprate are strong coupling superconductors.

Based on the above calculations as well as $T$c and isotope effect [31], a guiding principle toward higher $T$c can be suggested for cuprates: (1) Break the mirror symmetry, for example, by replacing $CuO_6$ octahedrons with $CuO_5$ pyramids to include the in-plane modes. (2) Every $CuO_2$ plane should accommodate doped holes equally [32]. (3) Then, there appear potentially a large number of hole carriers which interact with all phonon modes, contributing the Cooper pair formation. (4) Make the highest #1 band flatter to minimize the Fermi velocity and maximize the time to traverse the spin-correlated region, relative to the spin-flip time at the boundary. As a result, hole carriers can itinerate over a long distance without spin-scatterings. This guiding principle is consistent with the prediction that $T$c of cuprates increases with increasing the mean-free-path $\ell_o$ in a $CuO_2$ plane [30].

In summary, the non-rigid band theory of the electronic structure caused the phase change from the spin-disordered insulating state to the d-wave superconducting state. The FS structures changed from the coexistence of the Fermi pockets and the Fermi arcs in the underdoped regime to the Fermi arcs only in the overdoped regime. The origin of pseudogap was clarified.

## Acknowledgments

We would like to thank Profs. Atsushi Fujimori, Teppei Yoshida and Kazuyoshi Yamada for their valuable discussion on experimental results and Prof. Dung Hai Lee, Prof. Akira Shimizu and Dr. Franco Nori for helpful advice and encouragement regarding the present work. This work was supported by Tokyo University of Science and the Japanese Cabinet-Office-Projects: FIRST and ImPACT.